%% file: Savage_CIPANP_2015_v1p2.tex
\documentclass[12pt]{article}
\usepackage{graphicx}
\usepackage{color}

\usepackage[textwidth=6in]{geometry}

\definecolor{orange(ryb)}{rgb}{0.98,0.6,0.01}
\definecolor{orange(colorwheel)}{rgb}{1.0,0.5,0.0}
\definecolor{darkpowderblue}{rgb}{0.0,0.2,0.6}
\definecolor{forestgreen(web)}{rgb}{0.13,0.55,0.13}

\def\si{^1 \hskip -0.04in S _0}

\newcommand{\gsim}{\lower.7ex\hbox{$\;\stackrel{\textstyle>}{\sim}\;$}}
\newcommand{\lsim}{\lower.7ex\hbox{$\;\stackrel{\textstyle<}{\sim}\;$}}

\def\pislash{ {\pi\hskip-0.6em /} }

\def\nopi{ {\rm EFT}(\pislash) }

\newcommand\pubnumber{INT-15-057}
\newcommand\pubdate{\today}

\def\int{Institute for Nuclear Theory, University of Washington, Seattle, WA 98195-1550, USA.}
\def\support{\footnote{
Work supported in part by the Department of Energy under grant number 
DOE grant No.~DE-FG02-00ER41132}
}

\def\Title#1{\begin{center} {\Large #1 } \end{center}}
\def\Author#1{\begin{center}{ \sc #1} \end{center}}
\def\Address#1{\begin{center}{ \it #1} \end{center}}

\newcommand\pubblock{\rightline{\begin{tabular}{l} \pubnumber\\
         \pubdate  \end{tabular}}}
\newenvironment{Abstract}{\begin{quotation}  }{\end{quotation}}
\newenvironment{Presented}{\begin{quotation} \begin{center} 
             PRESENTED AT\end{center}\bigskip 
      \begin{center}\begin{large}}{\end{large}\end{center} \end{quotation}}
\def\Acknowledgements{\bigskip  \bigskip \begin{center} \begin{large}
             \bf ACKNOWLEDGEMENTS \end{large}\end{center}}

\input econfmacros.tex

\begin{document}
\begin{titlepage}
\pubblock

\vfill
\Title{Nuclear Physics from Lattice Quantum Chromodynamics}
\vfill
\Author{ Martin J. Savage\support}
\Address{\int}
\vfill
\begin{Abstract}
Quantum Chromodynamics  and Quantum Electrodynamics, 
both renormalizable quantum field
theories with a small number of precisely constrained input parameters, 
dominate the dynamics of the quarks and gluons - the underlying 
building blocks of protons, neutrons, and nuclei.  
While the analytic techniques of quantum field theory have played a key
role in understanding the dynamics of matter in high energy processes, they
encounter difficulties when applied to low-energy nuclear structure and
reactions, and dense systems.
Expected increases in computational resources into the exascale 
during the next decade will provide the ability to determine 
a range of important strong interaction processes directly from QCD
using the numerical technique of Lattice QCD.
This will complement the nuclear physics experimental program,
and in partnership with new thrusts in nuclear many-body theory, 
will enable unprecedented understanding and refinement of 
nuclear forces and, more generally, the visible matter in our universe.  
In this presentation, I will discuss the state-of-the-art Lattice QCD 
calculations of quantities of interest in nuclear physics,
progress that is expected in the near future, and the anticipated impact.
\end{Abstract}
\vfill
\begin{Presented}
Twelfth Conference on the Intersections of Particle and Nuclear Physics\\
Vail, Colorado, USA, May 19--24, 2015.
\end{Presented}
\vfill
\end{titlepage}
\def\thefootnote{\fnsymbol{footnote}}
\setcounter{footnote}{0}
%

\section{Introduction}

Lattice quantum chromodynamics (LQCD) has emerged from an extended period of research and development, 
dating back to the early 1970s~\cite{Fritzsch:1973pi,  Politzer:1973fx,Politzer:1974fr,Gross:1973id,Wilson:1974sk},
to now be in a position to calculate the properties and interactions of simple nuclear systems at the 
physical light quark masses with the inclusion of dynamical quantum electrodynamics (QED).
Hand-in-hand with exponentially increasing computational resources available to the fields of nuclear and particle physics, 
this progress has required  comparable algorithmic developments related to the discretization and evaluation of the 
quantum chromodynamics (QCD) path integral 
and also  theoretical developments in finite-volume and finite-lattice-spacing field theories and many-body theories.
With the anticipated growth in high-performance computing (HPC) resources available for such calculations, 
along with the continued support of the scientists that are critical to the success of this endeavor, we are entering a golden era of nuclear theory in 
which the nuclear forces will be refined directly from QCD, and reliable predictions for 
non-perturbative strong interaction quantities beyond the reach of  laboratory  investigation
will become possible.

Nuclei are comprised of neutrons and protons that essentially retain their identity within the nucleus and 
whose low-energy dynamics can 
be described by non-relativistic quantum mechanics.
In contrast,  nucleons  are entangled states 
of indefinite particle number comprised of massless gluons and nearly massless quarks.  
The  relativistic dynamics of the quarks and gluons are dictated by QCD, 
and quantum fluctuations play a central role,
through infrared slavery and spontaneous chiral symmetry breaking, 
in determining the emerging  properties and structure of the nucleon.
Remarkably, 
all of the diverse low-energy 
phenomenology of nuclei, from the magic numbers, through fission cross sections, through collective excitations and so forth, 
along with the properties and structure of the nucleons themselves, 
are  determined  (to relatively high precision)
by five fundamental parameters - the scale of the strong interactions, $\Lambda_{\rm QCD}$
the three light-quark masses, $m_u, m_d$ and $m_s$,
 and the fine-structure constant, $\alpha_e$.  
Upon reflection, it is surprising that nucleons retain so much of their ``identity''  in a nucleus 
rather than merging into one structureless ``blob'' of quarks and gluons.  
One aspect that can be addressed with the emerging LQCD technology is
the range of light-quark masses over which nuclei do, in fact, behave as collections of interacting nucleons.

It is important to appreciate how the  LQCD program is presently
impacting both the experimental and theoretical programs in nuclear physics. 
As LQCD is the only reliable technique with which to solve QCD in the low-energy regime, 
it is 
becoming increasingly important in designing and analyzing experiments.  
It makes contributions directly, for instance, in comparing the results of experiment with LQCD predictions, 
and
indirectly by providing key 
 inputs into other theoretical calculations that are relevant or vital for extracting physics.
In the US, the reach is extensive, impacting all four major areas of research in nuclear physics - 
hadron structure and fundamental symmetries, nuclear structure  and astrophysics, and hadrons spectroscopy.

The LQCD effort in the US is organized, in a significant way, by the USQCD collaboration~\cite{usqcd} 
of essentially all of the lattice QCD scientists in the US.
USQCD coordinates the production of ensembles of gauge-field configurations using computational resources acquired on 
leadership-class capability computing resources through peer-reviewed proposals, 
such as those submitted to the DOE's INCITE program.  
It also organizes the development and deployment  of 
LQCD algorithms and software through support from the DOE's SciDAC program~\cite{scidac} 
and collaboration with the SciDAC institutes.  
USQCD also operates its own capacity computing resources, such as CPU- and GPU-clusters, 
through DOE funding.  
This capacity computing continues to be  essential to the US's LQCD program.
Through its internal review process, proposal for scientific projects are supported with 
resources to perform computations in a coordinated manor to address  scientific goals of the nation.

\section{Methodology Overview}

The objective of the LQCD program 
is to make predictions for non-perturbative strong interaction observables in 
Minkowski space in infinite volume and in the continuum, while
LQCD calculations are performed in Euclidean space in a finite volume and with a discretized spacetime.
Consequently, an individual LQCD calculation does not directly calculate  predictions of QCD with arbitrary precision.
However, if the lattice spacing, $a$, is small compared with the scale of strong interactions, 
$a \ll \Lambda_\chi^{-1}$,
and if the volume, $L^3\times T$,
is large compared to the range of strong interactions, 
$m_\pi L, m_\pi T \gg 2\pi$,
then through the use of EFTs, extrapolations can be performed to make predictions for Euclidean-space QCD quantities.
For some quantities, such as hadron masses, this is the same as the Minkowski-space quantity, while for others, such as scattering amplitudes, further formal relations are required in order to translate these predictions into Minkowski space.
In addition, the QCD path integral requires summing over all values of all fields at all points in spacetime, 
and this is simply not possible through numerical  evaluation, and therefore sophisticated sampling techniques have to be employed.

A  LQCD result typically involves two or three distinct types of calculations that each require 
HPC resources.
The first is the production of the quantum fluctuations in the gluon fields that are 
stored as ensembles of gauge-field configurations.  
The production of such ensembles requires the largest capability 
supercomputers available, typically employing 
variants of the HMC algorithm to produce an extended  Markov chain of gluon-field configurations.
Fluctuations in the quark fields, the quark sea, are included implicitly through weightings with the determinant of the Dirac operator. 
Given the large human and computational resources required to create these configurations, they are often used for multiple observables,  both within collaborations and also by multiple collaborations through
agreements.

The second component of LQCD calculations is the generation of quark propagators from one spacetime point to another within the lattice volume of a given configuration of gluon fields.  
Such calculations  require significant capacity computing resources, and are performed on USQCD managed hardware or at NERSC, on XSEDE machines or with local university resources.
While  multiple techniques are available for such calculations, they all involve an iterative solution to produce a
unique propagator for the configuration and source structure.  
Recent algorithmic advances, such as multigrid and all-mode-averaging, have significantly reduced the resources required to calculate any given quark propagator.
As quark propagators are typically an order of magnitude larger in size
than the gauge-field configuration on which they are generated, 
and many are generated per configuration, it is relatively impractical to save them, and generally they are generated, used to construct correlation functions and then deleted.
An interesting aspect of the quark propagators is that  pion, nucleon, $\Delta$, deuteron, $^3$He and  $^4$He 
 correlation functions
can all be generated (in the limit of exact isospin symmetry)
from a single light-quark propagator. 
Therefore, in the ensemble average, 
lengths scales associated with the pion must  cancel in forming the correlation functions for the $A=1,2,3,4$ nuclei.  
It is the statistical nature of this cancellation due to the finite sampling of the QCD path integral that 
is responsible for the exponential signal-to-noise degradation  exhibited by such 
correlation functions at late times.

Finally,  
hadronic
correlation functions are constructed in order to access the physics of interest, be it the mass of a hadron or a matrix element.
The construction of the appropriate sources and sinks with which to calculate 
correlation functions is somewhat of an art as one only has an approximate idea of 
what the actual QCD solution is for the system under exploration.
If the sources and sinks do not resemble, at some level, the system of interest, then the overlap onto it will be small 
and it will not be possible to extract that state from the  correlation function. 
 On the other hand,  if they strongly resemble the state of interest,  it will dominate the correlation function.  
 Of course, one desires   the second scenario.
In reality, multiple sources and sinks are used to generate a matrix of correlation functions, from which the lowest-lying 
eigenstates in the lattice volume are identified.

\section{ The Physics Program }

\begin{figure}[ht!]
\centering
\includegraphics[height=2.5 in]{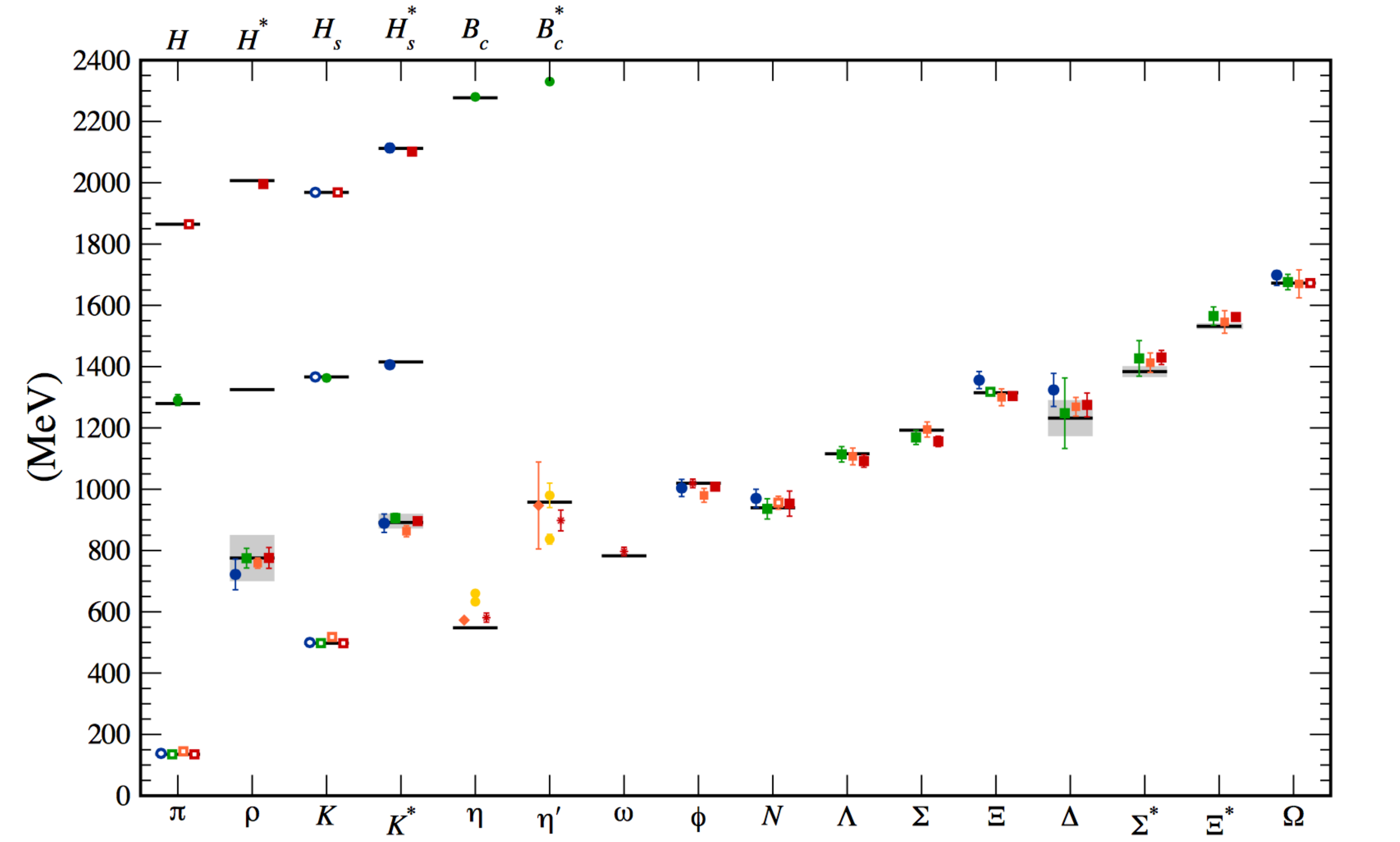}
\caption{
A summary of light hadron masses postdicted by LQCD~\protect\cite{Kronfeld:2012uk}.
The colors denote distinct sets of calculations and open points indicate the hadron was used to tune the input parameters. 
[Figure reproduced with permission from Andreas Kronfeld.]
}
\label{fig:Masses}
\end{figure}

The light quark masses and the scale of the strong interactions are constrained to
reproduce a small number of select meson and baryon masses,
and now provide the most precise constraints on the values of the light-quark masses.
Tuning these quantities  
is an iterative procedure,  and interpolations or extrapolations 
of observables to the physical point are generally required.
As an example, in the isospin limit and without QED, the 
 pion, kaon and $\Omega^-$ masses can be used to fix the input parameters of a LQCD calculation.
 In most cases, it would be ideal to work at the physical values of the parameters, but the computational resources required for such calculations 
 are only now becoming available. 
There are   number of LQCD calculations of the light and heavy hadron spectra that can be directly compared with experiment, 
see Fig.~\ref{fig:Masses}.
LQCD successfully postdicts the spectra of light and heavy hadrons, accomplishing an important verification step.

\begin{figure}[ht!]
\centering
\includegraphics[height=2.1 in]{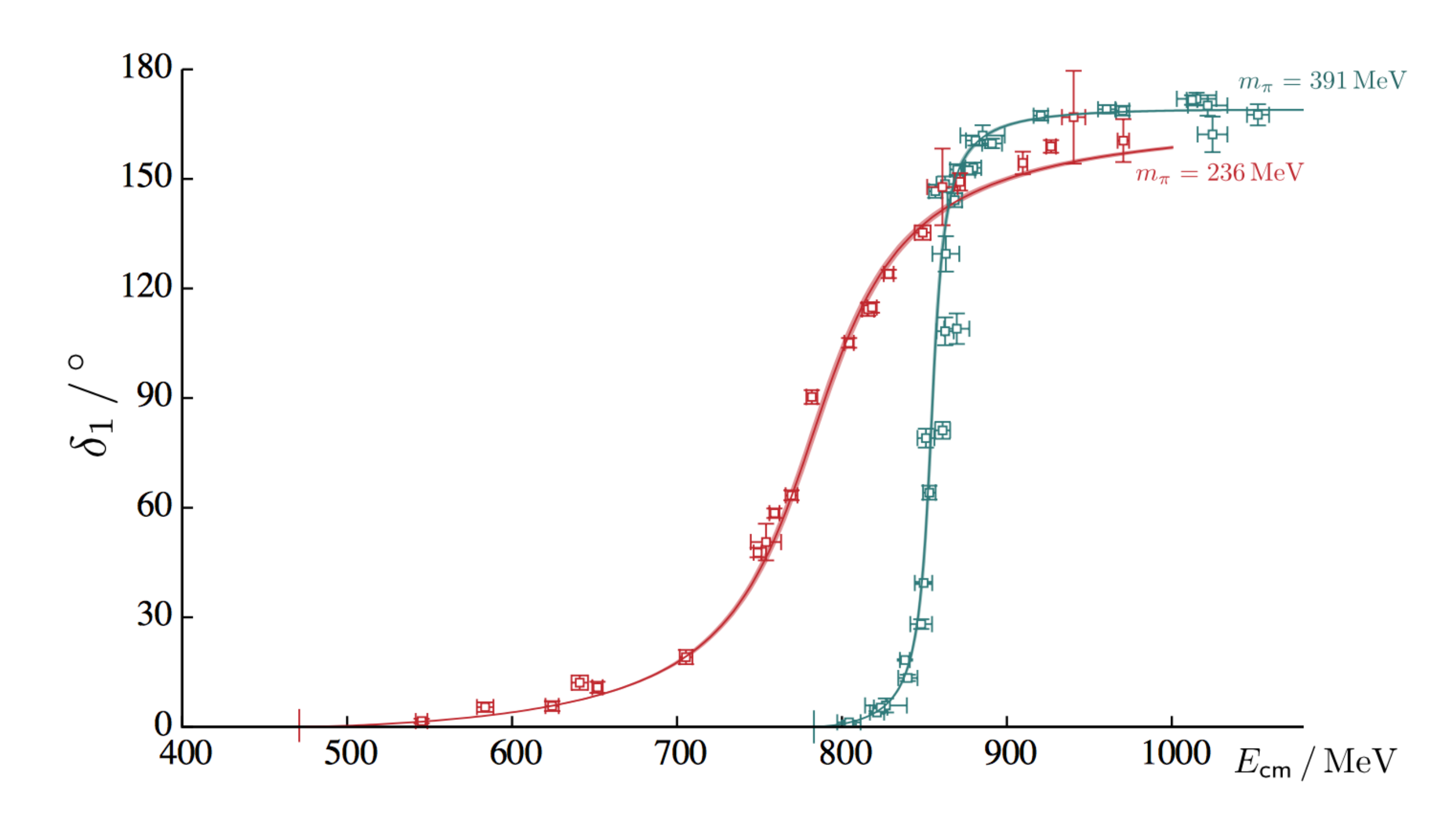}\ \ \includegraphics[height=2.1 in]{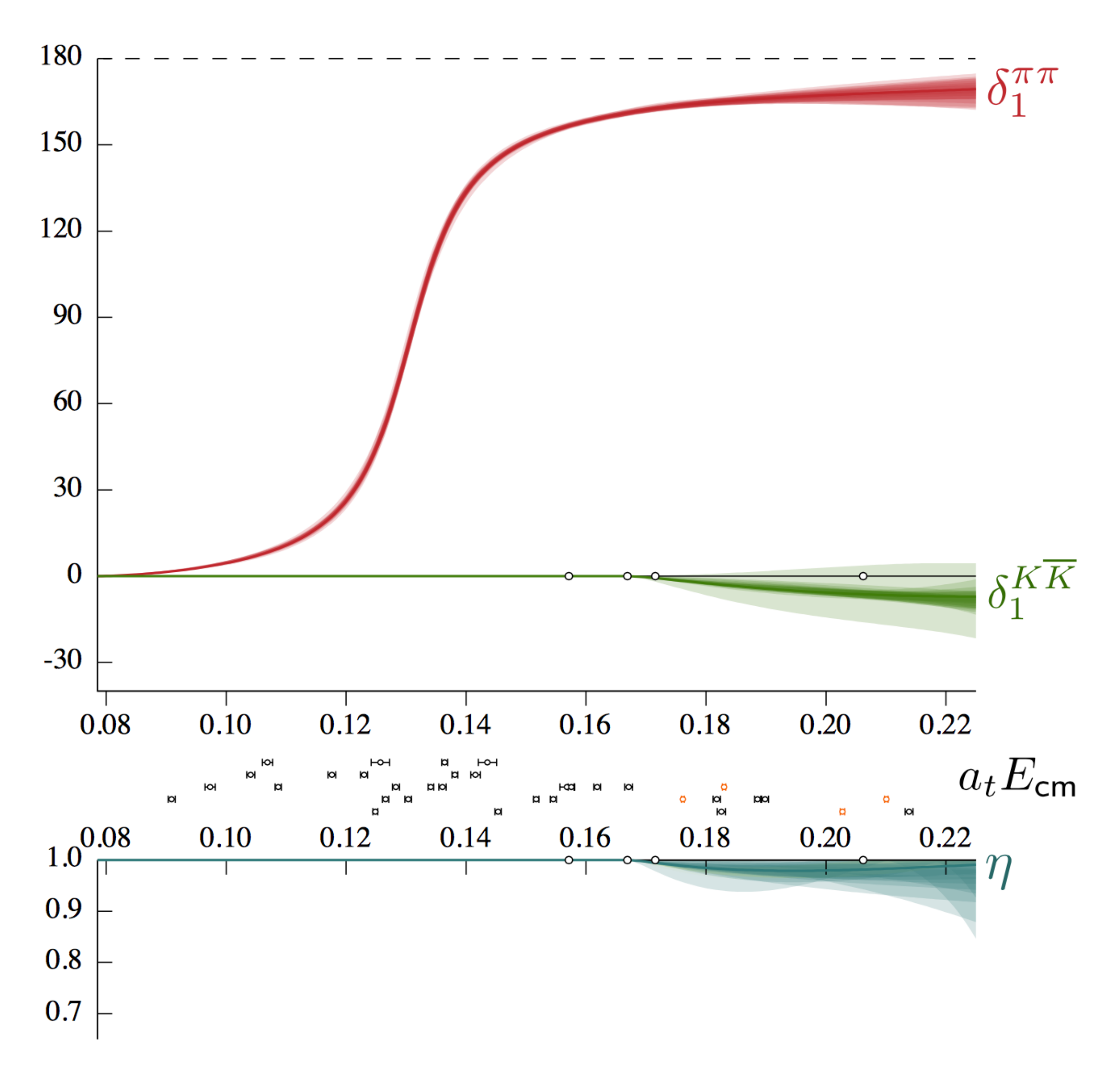}
\caption{
The left panel shows the $\rho$ resonance calculated with 
$m_\pi\sim 240~{\rm MeV}$ and  $m_\pi\sim 390~{\rm MeV}$~\protect\cite{Wilson:2015dqa}.  
The right panel shows the parameters of the two-channel S-matrix describing the 
$\pi\pi$-$K\overline{K}$ scattering in the $\rho$ channel
around the inelastic threshold at $m_\pi\sim 240~{\rm MeV}$~\protect\cite{Wilson:2015dqa}.
[Figures reproduced with permission from Robert Edwards.]
}
\label{fig:Jlab}
\end{figure}

One of the major efforts underway in the US nuclear physics LQCD community is to  determine the 
mesonic and baryonic excitation spectra.   
While there are a number of good reasons for investing in such calculations, one of the more important is to discover exotic states in the spectra that lie beyond description in simple hadronic models, and directly probe the field theory ingredients of QCD.  
The Hadron Spectrum collaboration, centered around JLab,  is leading this effort in the US and 
continues to produce pioneering and exciting results that are playing a key role in the 
GlueX experiment~\cite{gluex} at JLab, e.g. Refs.~\protect\cite{Wilson:2015dqa,Wilson:2014cna,Morningstar:2013bda}.  
Their calculations are able to resolve multiple states with the same quantum numbers, both those with ``ordinary'' quantum numbers and ``exotic'', and they are able to determine aspects of 
the composition and structure of the state through the strength of coupling to a range of source and sink structures.
A critical and challenging aspect of this program is to be able to describe the spectra in the presence of 
multi-channel and multi-body states which become more dense as the pion mass becomes lighter.
This has required a concerted effort to develop  finite-volume formalism that is at the heart of such an analysis~\cite{Briceno:2012yi}. 
Recently, there have been a number of results demonstrating that the two-body coupled channels systems can be analyzed in a systematic  
way with a complete quantification of associated uncertainties.
The $\rho$-resonance has been convincingly mapped out at different pion masses, 
as shown in Fig.~\ref{fig:Jlab}.
%


The structure of the nucleon remains a forefront area of research in nuclear physics. 
While JLab and RHIC provide precise probes of the quark structure of the nucleon and nucleus, 
a precise
mapping of their gluon structure will have to wait for an operational electron-ion collider (EIC).
While the structure of the nucleon remains a somewhat confusing and yet fascinating subject, 
particularly when the dependence on the quark masses is considered, 
the conflicting experimental measurements of the proton charge radius is  disturbing given 
that it is such a basic quantity.  The charge radius measured with muonic-hydrogen is significantly smaller than that measured with electrons.  
So far, there are no satisfactory explanations that reconcile these results, however, reinvestigations of the radiative corrections to the electronic measurements may yield something interesting~\cite{Lee:2015jqa}.
With LQCD calculations now being performed at the physical quark masses and with ability to include QED, 
LQCD is in a position, with access to sufficient resources,
 to calculate the nucleon charge radius with sufficient  precision to 
distinguish between the two experimental values.


When the idea that the strange quark may play a nontrivial role in the structure of the nucleon was raised, 
launching an extensive experimental program that has lasted decades, 
it was  considered odd  in terms of the hadronic models of the 1960's, 
but considered natural in the context of QCD.
Given that strong interaction properties are generally of ``order one'', it was therefore natural that 
matrix elements of strange-quark operators should also be of order one, 
unless symmetry arguments required otherwise.  
The enormous experimental effort that followed has shown that strange-quark observables are much smaller than one naively expects, rendering them somewhat ``uninteresting'' (except for their smallness!).
With the computational and algorithmic developments in LQCD, 
calculations of disconnected diagrams, i.e. those which will generate non-zero strange-quark matrix elements,
are now possible with  precision.  
For the simple observables that have been calculated, all strange-quark matrix elements have been found to be small, 
consistent with what has been observed experimentally.
\begin{figure}[ht!]
\centering
\includegraphics[height=2.8 in]{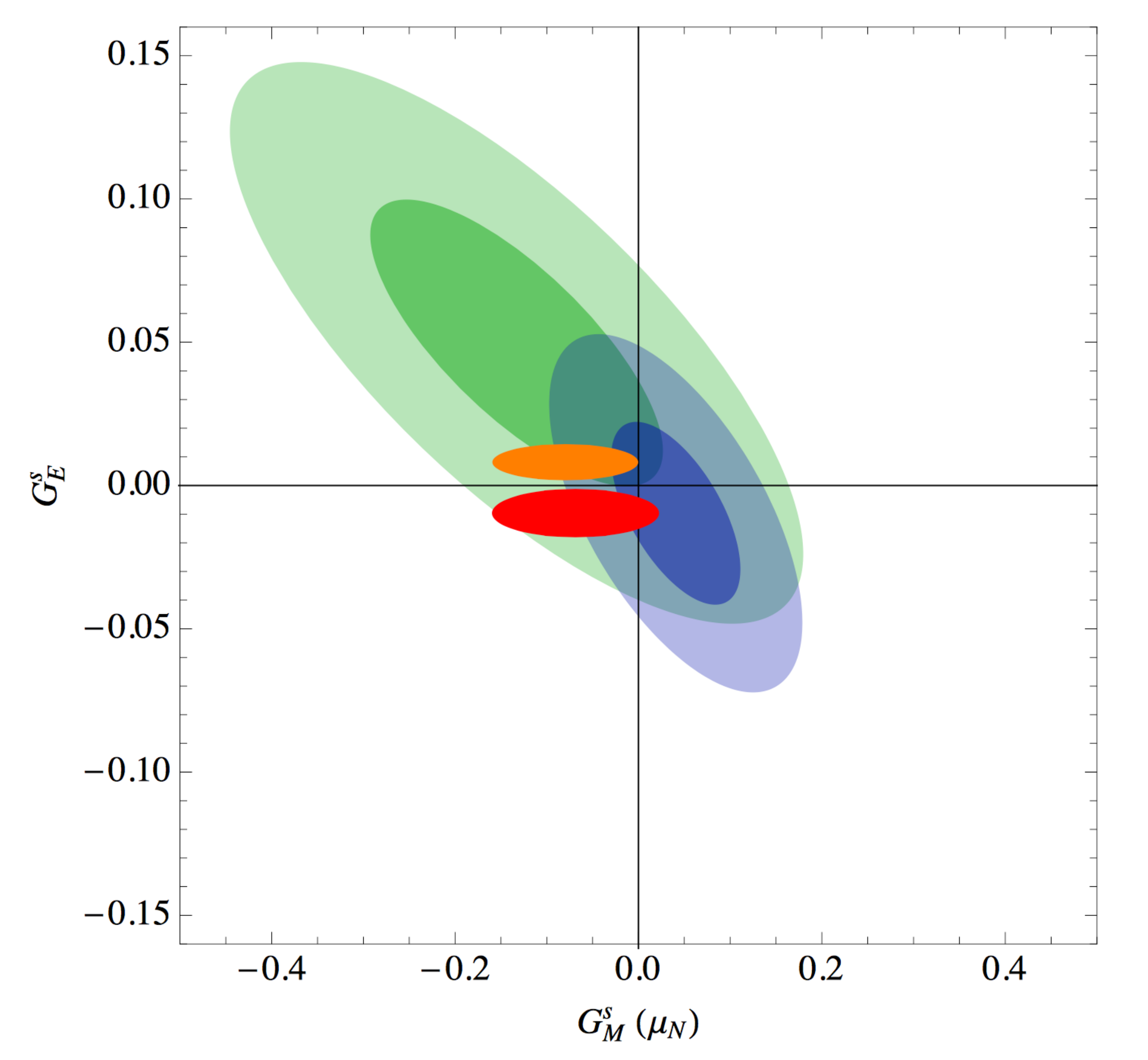}\caption{
The strange form factors $G_E^{(s)}$ and $G_M^{(s)}$~\protect\cite{Shanahan:2014tja}.
The 
\textcolor{red}{red} and \textcolor{orange(colorwheel)}{orange} 
ellipses are the results of LQCD calculations at 
$Q^2=0.26~{\rm GeV}^2$ and $Q^2=0.17~{\rm GeV}^2$, respectively, 
while the 
\textcolor{darkpowderblue}{blue}  and \textcolor{forestgreen(web)}{green} 
ellipses show the experimental constraints 
at $Q^2=0.23~{\rm GeV}^2$.
[Figure reproduced with permission from Phiala Shanahan.]
}
\label{fig:GEGM}
\end{figure}
Figure~\ref{fig:GEGM}
shows the strange form factors $G_E^{(s)}$ and $G_M^{(s)}$ evaluated at comparable values of $Q^2$~\cite{Shanahan:2014tja}.
The smaller ellipses denote the results of LQCD calculations and should be compared with the larger ellipses resulting from experimental measurements.
It must now be concluded that  strange-quark matrix elements in the nucleon 
are more precisely known from LQCD calculations than they are from experiment.

\begin{figure}[ht!]
\centering
\includegraphics[height=3.5 in]{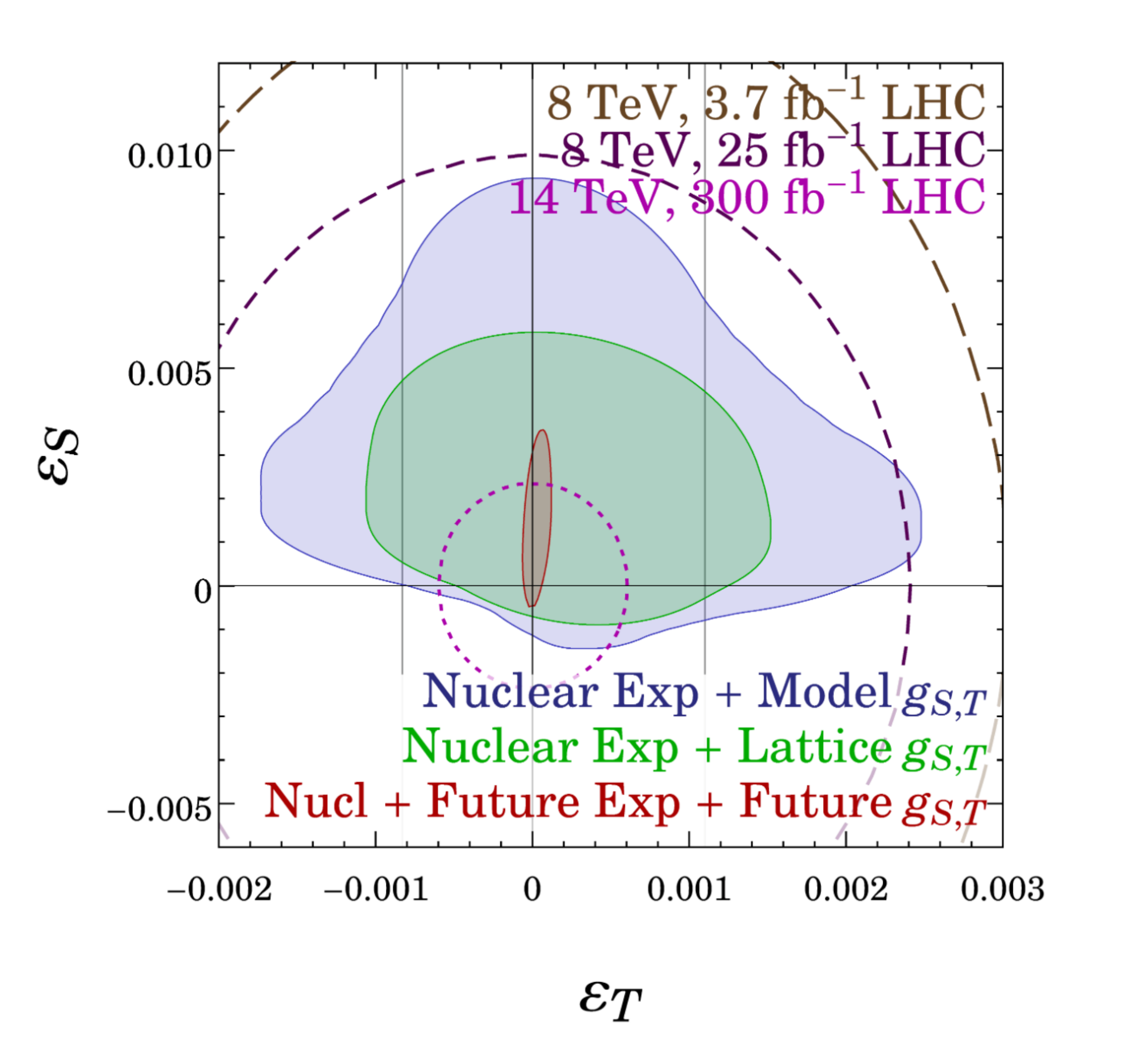} 
\caption{
Constraints on the scalar, $\epsilon_S$, and tensor,  $\epsilon_T$, matrix elements in the nucleon from ongoing nuclear physics 
experiments combined with LQCD calculated nucleon matrix elements, 
along with present and expected constraints from the LHC~\cite{Gupta:2014dla}.
[Figure reproduced with permission from Rajan Gupta.]
}
\label{fig:poundme}
\end{figure}

Tests of the standard model through precision measurements
of the properties and decays of nucleons and nuclei require knowledge of 
strong interaction matrix elements.
The $\beta$-decay of the neutron has long been used as a process with which to search for new physics.
First, by establishing, in part,
 the chiral nature of the weak interactions through the 
charged-current
$V-A$ couplings to the electron 
and antineutrino, and now as away to search for physics beyond the standard model (BSM). 
New precision experiments at LANL measuring  the neutron lifetime, angular distributions and spin correlations, 
along with precise LQCD calculations of the matrix elements of quark bilinears in the nucleon, are permitting tight 
constraints to be placed 
on  non-standard model couplings between the quarks and 
leptons~\cite{Bhattacharya:2011qm,Gupta:2014dla,Bhattacharya:2015wna}.  
These are expected to be significantly more constraining than 
those that can be obtained at the LHC during the next several years,
as shown in Fig.~\ref{fig:poundme}.
%


One of the important theoretical developments of the 1990's was the establishment of the chiral nuclear forces.  
Until Weinberg's pioneering papers in the early 1990's~\cite{Weinberg:1990rz,Weinberg:1991um}, 
the approximate chiral symmetry of QCD had not been systematically included 
in the phenomenological potentials used to describe the forces between nucleons.
Weinberg, used the hierarchy of interactions defined in chiral perturbation theory ($\chi$PT) 
to define two-particle irreducible diagrams that constitute the chiral nuclear interactions~\cite{Ordonez:1992xp}. 
Divergences in the Lippman-Schwinger equation, along with the peculiarities of singular potentials, complicate the analysis of multi-nucleon systems
in nuclear EFTs (NEFTs)
far beyond what was initially imagined from the one-nucleon 
sector~\cite{Kaplan:1996xu,Kaplan:1998tg,Kaplan:1998we,Beane:2001bc}, which is comfortably described by Heavy-Baryon-$\chi$PT.  
In addition to providing a systematic expansion of the nucleon-nucleon interactions with a well-defined power counting of small expansion parameters, 
chiral forces also make clear the order at which multi-nucleon forces contribute.  
This expansion has now been carried out to a few orders in Weinberg's power-counting, 
and the agreement with experimental data is found to improve order-by-order~\cite{Epelbaum:2008ga,Machleidt:2011zz}.

A deficiency of NEFTs is that new counterterms are required at each order 
in the expansion.
It is this point that was the main reason driving the formation of the NPLQCD
collaboration.
The only way to continue to improve the precision of predictions from 
NEFTs is to be able to determine the necessary counterterms from QCD, 
and the only way to do this reliably is with LQCD.
The current plan is to use LQCD to calculate the low-lying spectra of   light nuclei in the  S-shell and some in the P-shell, and also nucleon-nucleon scattering phase shifts, and then, as with experimental data, constrain the form of the nuclear forces, which translates into determining counterterms in the chiral nuclear forces with some level of precision.
A complete dissection of the chiral forces requires the dependence on the light-quark masses to be isolated, and 
consequently LQCD calculations at and around the physical point are required to develop precise predictive capabilities in nuclear many-body systems using chiral nuclear forces.

\begin{figure}[ht!]
\centering
\includegraphics[height=2.6 in]{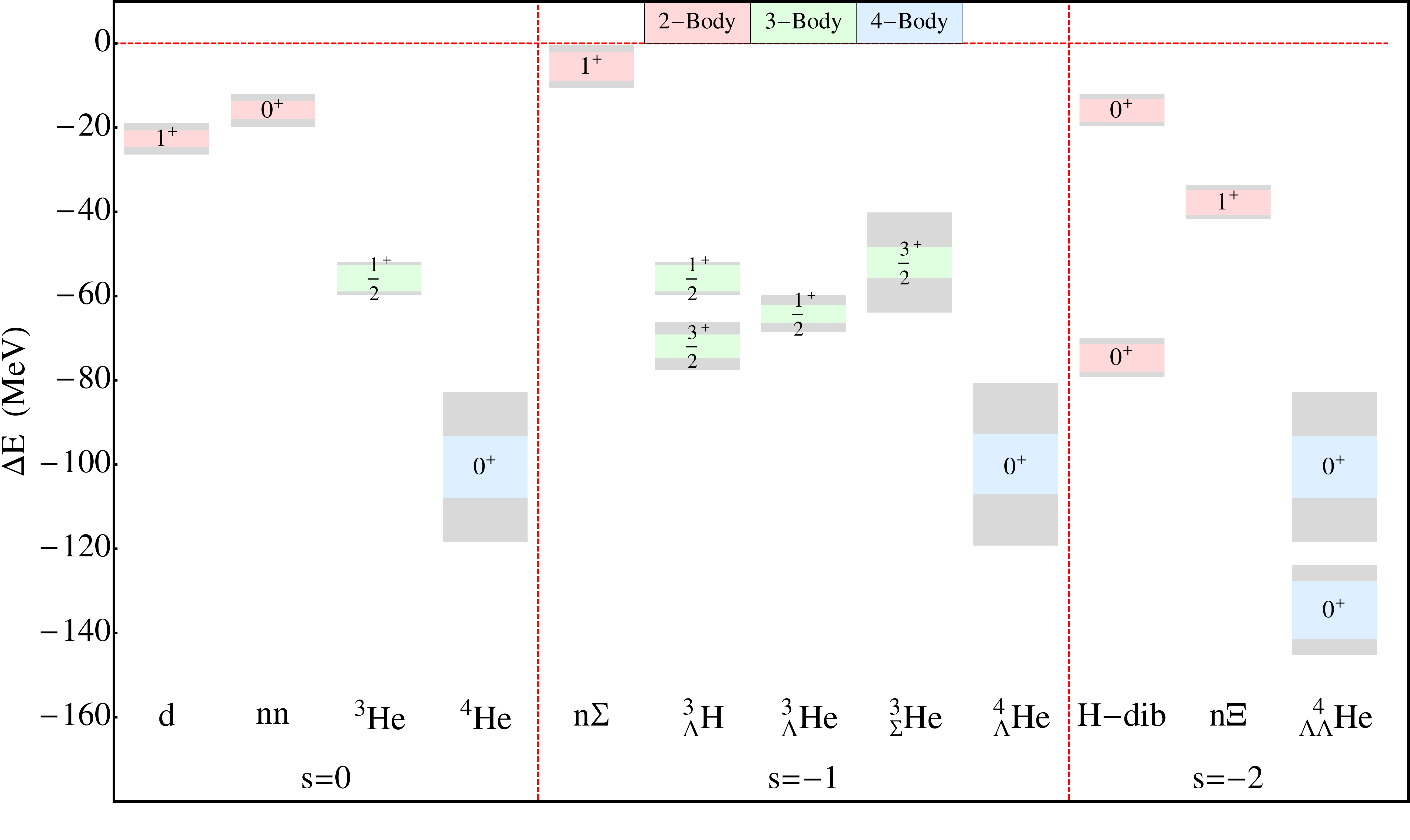} 
\caption{
The lowest-lying states in S-shell nuclei and hypernuclei at a pion mass of
$m_\pi\sim 805~{\rm MeV}$~\protect\cite{Beane:2012vq}.
  }
\label{fig:su3}
\end{figure}
\begin{figure}[ht!]
\centering
\includegraphics[height=1.85 in]{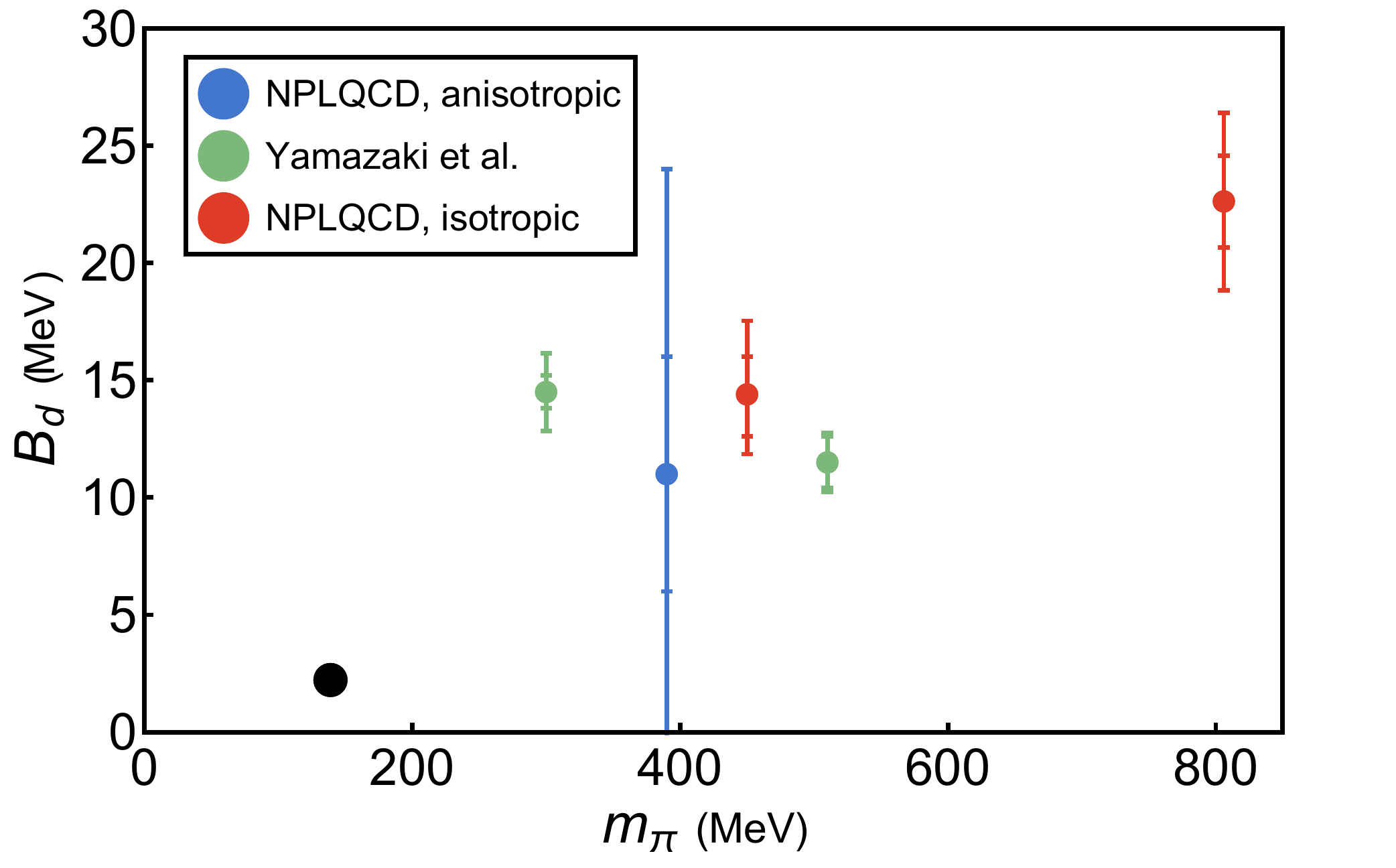} \includegraphics[height=1.9 in]{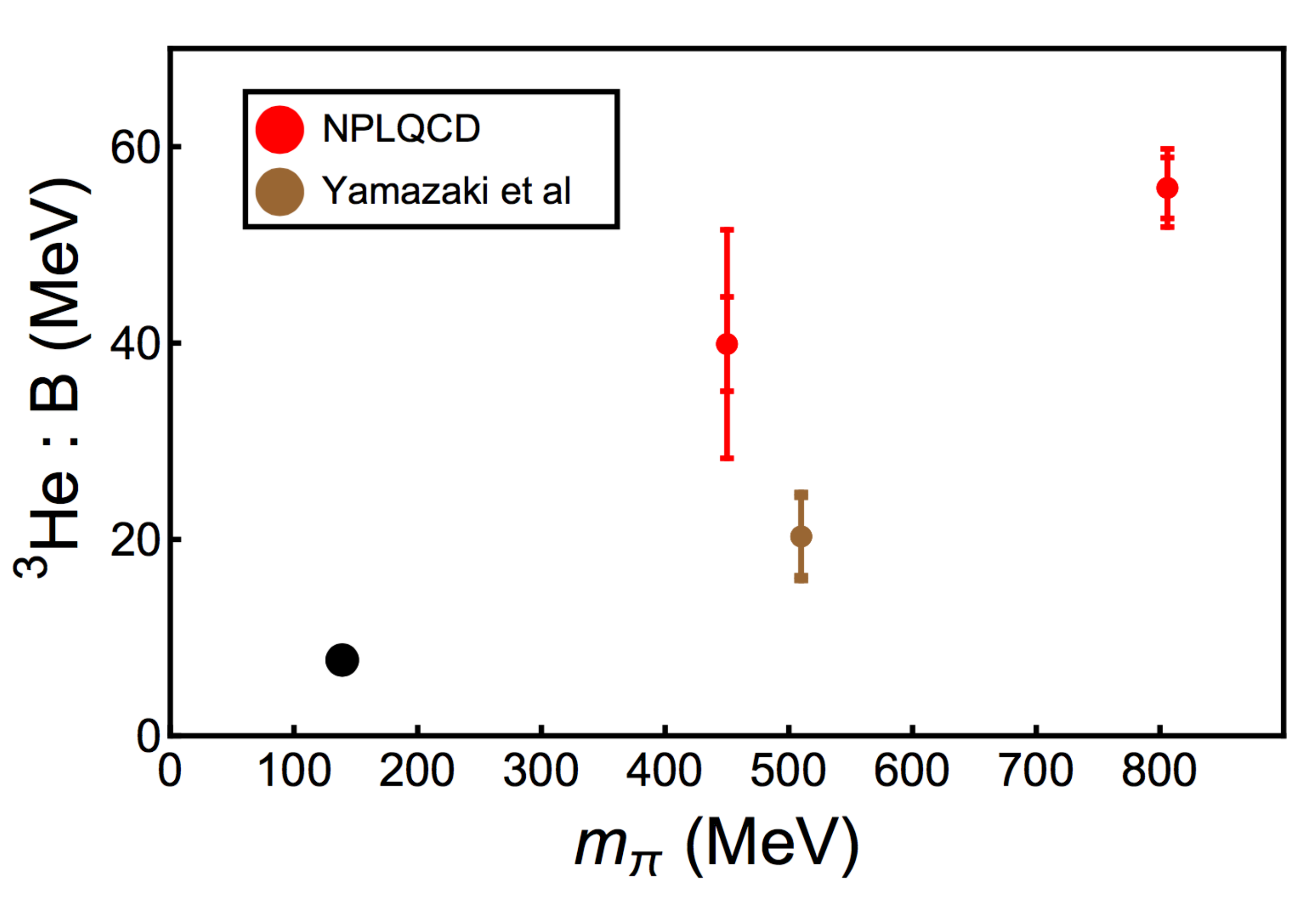} 
\caption{
The deuteron~\protect\cite{Orginos:2015aya} (left panel) and $^3$He (right panel) binding energies as a function of pion mass~\protect\cite{Beane:2012vq,Beane:2011iw,Yamazaki:2012hi,Yamazaki:2015asa}.
  }
\label{fig:DeutBind}
\end{figure}

Extensive calculations have been performed at a pion mass of $m_\pi\sim 805~{\rm MeV}$, 
the point where SU(3)-flavor symmetry is exact and the strange-quark mass is tuned to that of nature.  
The lowest-lying states in the S-shell nuclei and many hypernuclei have been 
calculated~\cite{Beane:2012vq}, see Fig.~\ref{fig:su3},
and the 
nucleon-nucleon phase shifts have been determined over a 
significant range of momenta~\cite{Beane:2013br}.
One of the interesting observations is that the scattering parameters, scattering length and effective range, suggest that the deuteron remains a``large and fluffy'' object over a range of pion masses, indicating that while it is unnatural it is not fine-tuned.  
The same cannot be said in the $\si$ channel, where the nucleon-nucleon system is both unnatural and fine-tuned.
The binding of the lightest nuclei, including the deuteron and $^3$He, have been determined at a number of pion masses, as shown in Fig.~\ref{fig:DeutBind}.
These calculations have demonstrated that  light nuclei become more deeply bound with increasing pion mass,
eliminating one of the  scenarios that was conjectured by NEFTs~\cite{Beane:2006mx} prior to LQCD calculations,
based upon naive dimensional analysis (NDA).

\begin{figure}[ht!]
\centering
\includegraphics[height=2.5 in]{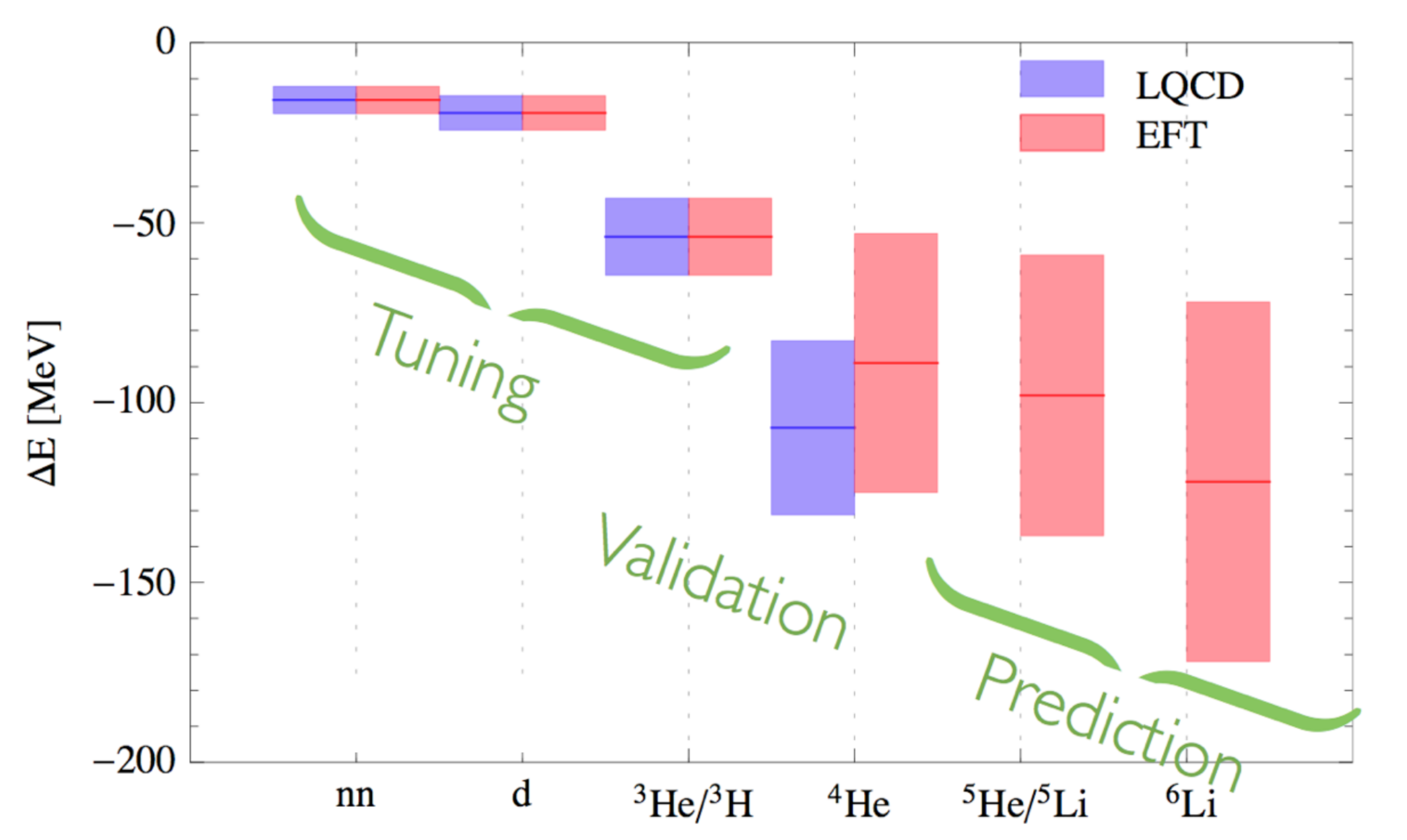} 
\caption{
Predictions for the binding energies of $^5$He, $^5$Li and $^6$Li~\protect\cite{Barnea:2013uqa} 
obtained from $\nopi$ tuned to LQCD 
calculations of the binding of the deuteron, dineutron and $^3$He~\protect\cite{Beane:2012vq,Beane:2013br}.
[Figure reproduced with permission from William Detmold.]
  }
\label{fig:pionlessPRED}
\end{figure}

One of the exciting developments of the last year or so has been the first matching to nuclear many-body calculations and  
subsequent predictions for the binding energies of nuclei with mass number beyond that of the LQCD calculations.  
More than a decade ago, 
the matching between LQCD calculations and NEFTs
was identified as a critical part of  the ``bridge'' between QCD and nuclei that is required for direct connection between 
quarks and gluons and the properties and interactions of larger nuclei, and now the bridge is starting to be built!  
In some beautiful work~\cite{Barnea:2013uqa},  the ground-state energies of the A=2,3,4 nuclei 
at $m_\pi\sim 805~{\rm MeV}$~\cite{Beane:2012vq} were used to determine the two-nucleon 
and three-nucleon interactions in the pionless EFT, 
$\nopi$~  \cite{Kaplan:1998we,vanKolck:1998bw,Chen:1999tn}, 
which is valid up the t-channel cut in NN scattering, 
$|{\bf p}| \lsim 400~{\rm MeV}$.
The subsequent prediction of the ground state of $^4$He was used to verify the 
matching (and LQCD prediction).  
Once verified, $\nopi$ was used to predict the binding energies of  $^5$He, 
$^5$Li and $^6$Li, 
as shown in Fig.~\ref{fig:pionlessPRED},
extending the reach of the LQCD calculations in a way that is consistent with the symmetries of QCD using the small expansion parameter defining $\nopi$.
While these calculations are only relevant to an unphysical universe, they do allow us to explore the periodic table  QCD with different light-quark masses.
Importantly, it demonstrates the path forward at the physical point, albeit with a different NEFT.
%


The quark-mass dependence of the nucleon mass can be used to determine the nucleon $\sigma$-term
via the Feynman-Hellman theorem, which is equal to matrix elements of 
$\sum\limits_q m_q \overline{q} q$.  
This construction straightforwardly extends to nuclei, 
and by forming ratios of the nuclear $\sigma$-terms to the nucleon $\sigma$-term, 
the modifications due to nuclear forces, explicitly the nuclear binding, 
can be determined~\cite{Beane:2013kca}. 
In the isospin limit, the light-quark mass cancels, leaving the ratio of isoscalar-scalar matrix elements, 
i.e. $ \langle \overline{q} q \rangle_A /  ( A \langle \overline{q} q \rangle_N )$, providing a measure of
the deviations from a collection of non-interacting nucleons.
This is important  in the context of Dark Matter detection by the recoil of nuclear targets, 
in particular, in calculating the nuclear response to such interactions at  the $\lsim 10\%$ level.
\begin{figure}[ht!]
\centering
\raisebox{0.25\height}{\includegraphics[height=1.25 in]{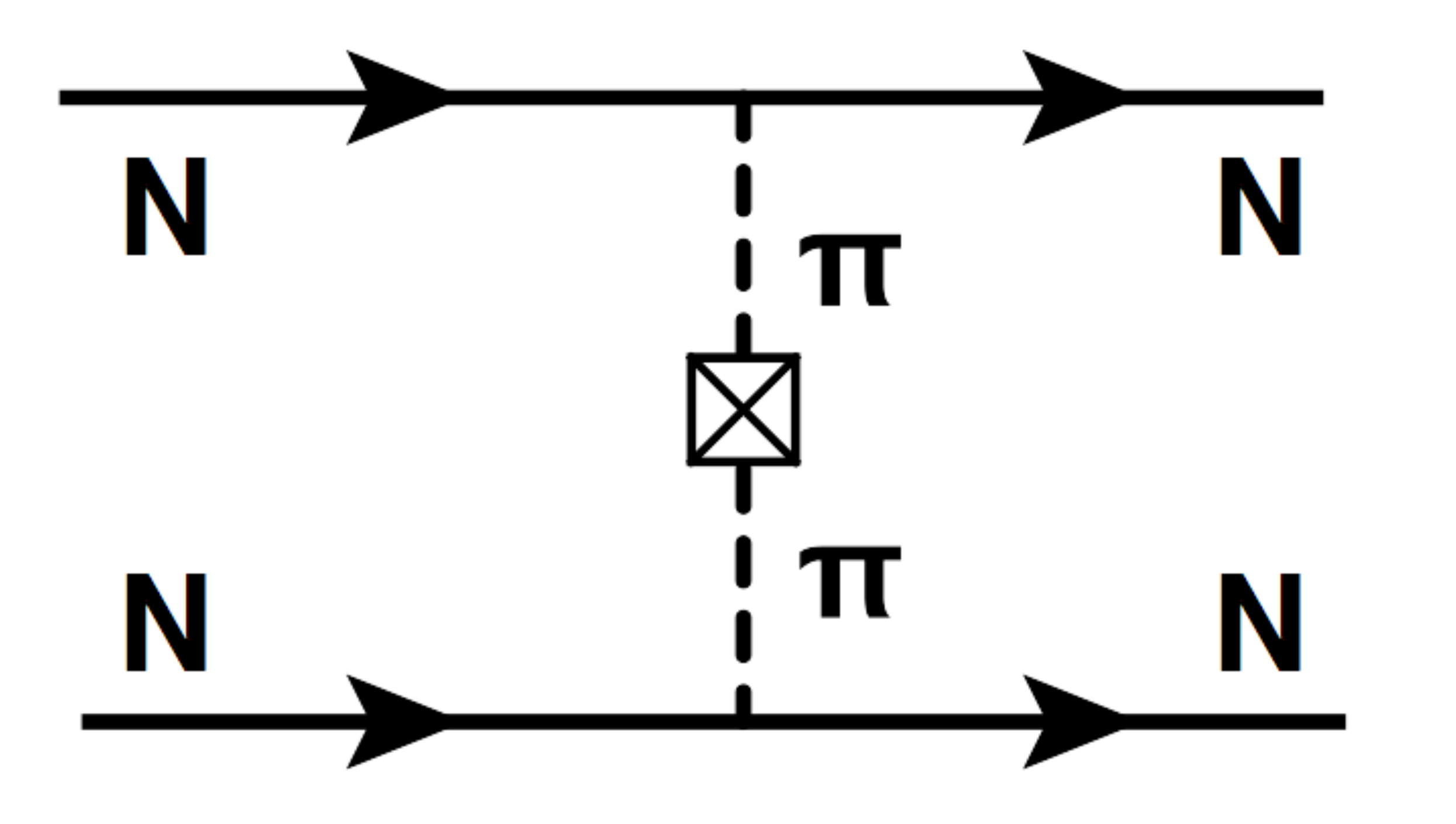} } \ \ \ \includegraphics[height=1.75 in]{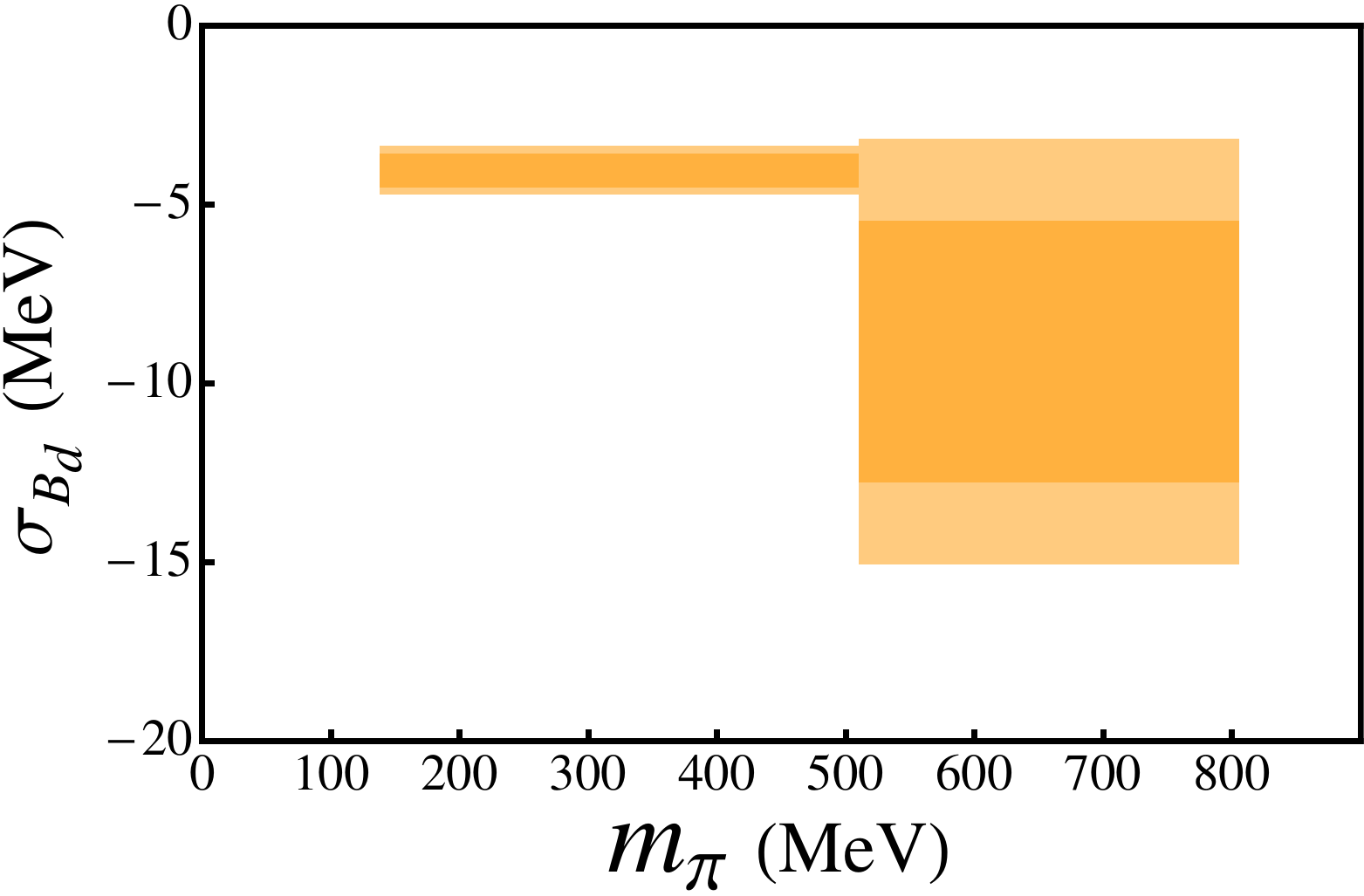} 
\caption{
The left panel shows one of the interactions that an isoscalar-scalar Dark Matter particle can have with nuclei that is beyond 
interactions with single nucleons. 
The right panel shows results of LQCD calculations of the contribution to the deuteron $\sigma$-term from 
nuclear interactions that are beyond single nucleon contributions from LQCD calculations~\protect\cite{Beane:2013kca}.
}
\label{fig:sigmaterm}
\end{figure}
%

\begin{figure}[ht!]
\centering
\includegraphics[width=0.6\columnwidth]{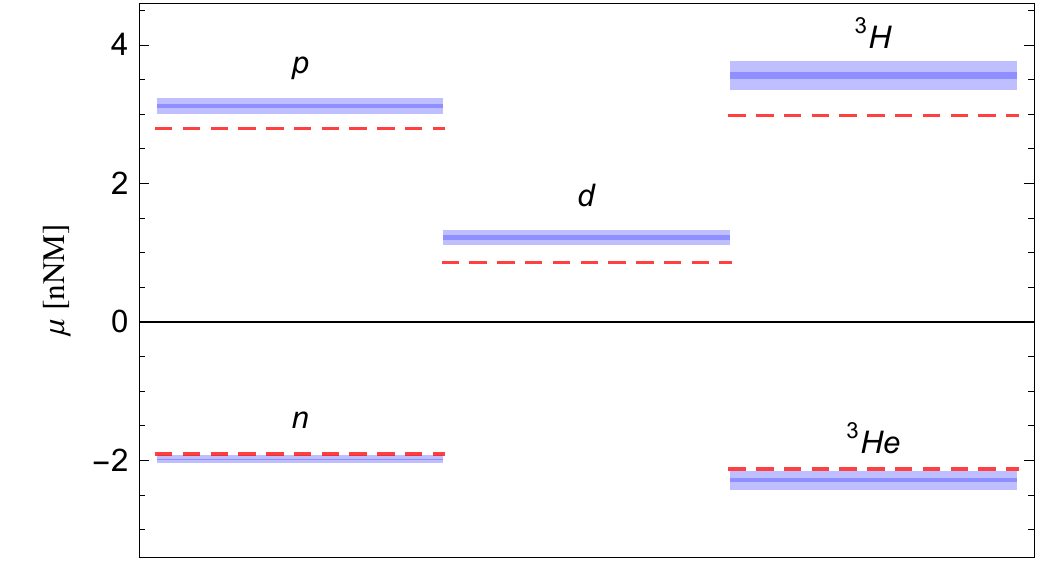}
\caption{
  The results of the Lattice QCD calculations of nuclear magnetic 
  moments~\protect \cite{Beane:2014ora,Chang:2015qxa}
 are shown as the blue bands, 
  while the corresponding experimental values are shown by the red dashed lines.
  Natural Nuclear Magnetons have been used, in which the nuclear magneton is defined in terms of the nucleon mass 
  at the given value of the light-quark masses.
}
\label{fig:magmoms}
\end{figure}

Recently, the magnetic moments of the lightest nuclei have been calculated at a pion mass of 
$m_\pi\sim 805~{\rm MeV}$ by performing LQCD calculations in the presence of a uniform and 
time-independent magnetic field.
A summary of the magnetic moments of the nucleon and lightest nuclei is shown in Fig.~\ref{fig:magmoms}.
It is remarkable that, when given in terms of natural Nuclear Magnetons (defined with the nucleon mass at the given light-quark masses), the magnetic moments of the light nuclei at a pion mass of 
$m_\pi\sim 805~{\rm MeV}$
are very close to their values at the physical light-quark masses. 
This implies that essentially all of the light-quark mass dependence of the magnetic moments is determined by the mass of the nucleon, and that a non-relativistic quark model type scenario is providing the dominant contribution - one in which a naive weighted sum of quark-model quark spins is dominant. 
This could  be a consequence of the large-Nc limit of QCD or it might be something more. 
Another, and perhaps more remarkable, feature is that the magnetic moments of the nuclei are essentially given by the sum of the nucleon magnetic moments in a naive nuclear shell-model configuration. 
For these nuclei, this is somewhat trivial compared to larger nuclei, but nonetheless the two neutrons in the triton
are largely in a spin-zero configuration with the magnetic momentum being essentially that of the proton. 
The deviations observed in nature from the naive shell-model prediction are in agreement, 
within uncertainties, with the result of the LQCD calculation at the heavier pion mass. 
This leads one to observe that nuclei behave as collections of weakly interacting nucleons over a large range of light-quark masses, and that the phenomenological nuclear shell-model is not limited in 
applicability to the physical point, but is somewhat of a generic feature of QCD for arbitrary 
light-quark masses. 
It will be interesting to learn if there are values of the quark masses for which nuclei collapse into a strongly interacting configuration of quarks and gluons rather than of weakly nucleons with a hierarchy of multi-nucleon forces.


Calculations of two-nucleon systems in background magnetic fields 
were used to isolate the short-distance two-body electromagnetic contributions to the radiative 
capture process $np\rightarrow d\gamma$, and the photo-disintegration processes 
$\gamma^{(*)}d\rightarrow np$~\cite{Beane:2015yha}. In nuclear potential models, such contributions are
described by phenomenological meson-exchange currents, while we were able to determine
them directly from the quark and gluon interactions of QCD. 
Calculations of neutron-proton
energy levels in multiple background magnetic fields were performed at two values of the
quark masses, corresponding to pion masses of 
$m_\pi\sim 450~{\rm MeV}$ and $m_\pi\sim 805~{\rm MeV}$, and were combined 
with $\nopi$ to determine this low-energy inelastic process. 
Extrapolating to
the physical pion mass, a cross section of 
$\sigma^{\rm lqcd}(np\rightarrow d\gamma)=334.9^{+5.4}_{-4.7}$ mb
was obtained at an incident neutron speed of 
$v=2200~{\rm m/s}$, consistent with the experimental value of 
$\sigma^{\rm expt}(np\rightarrow d\gamma)=334.2\pm 0.5$ mb. 
This is the first LQCD calculation of an inelastic nuclear reaction.

\begin{figure}[!ht]
  \centering  
 \includegraphics[height=2.5 in ]{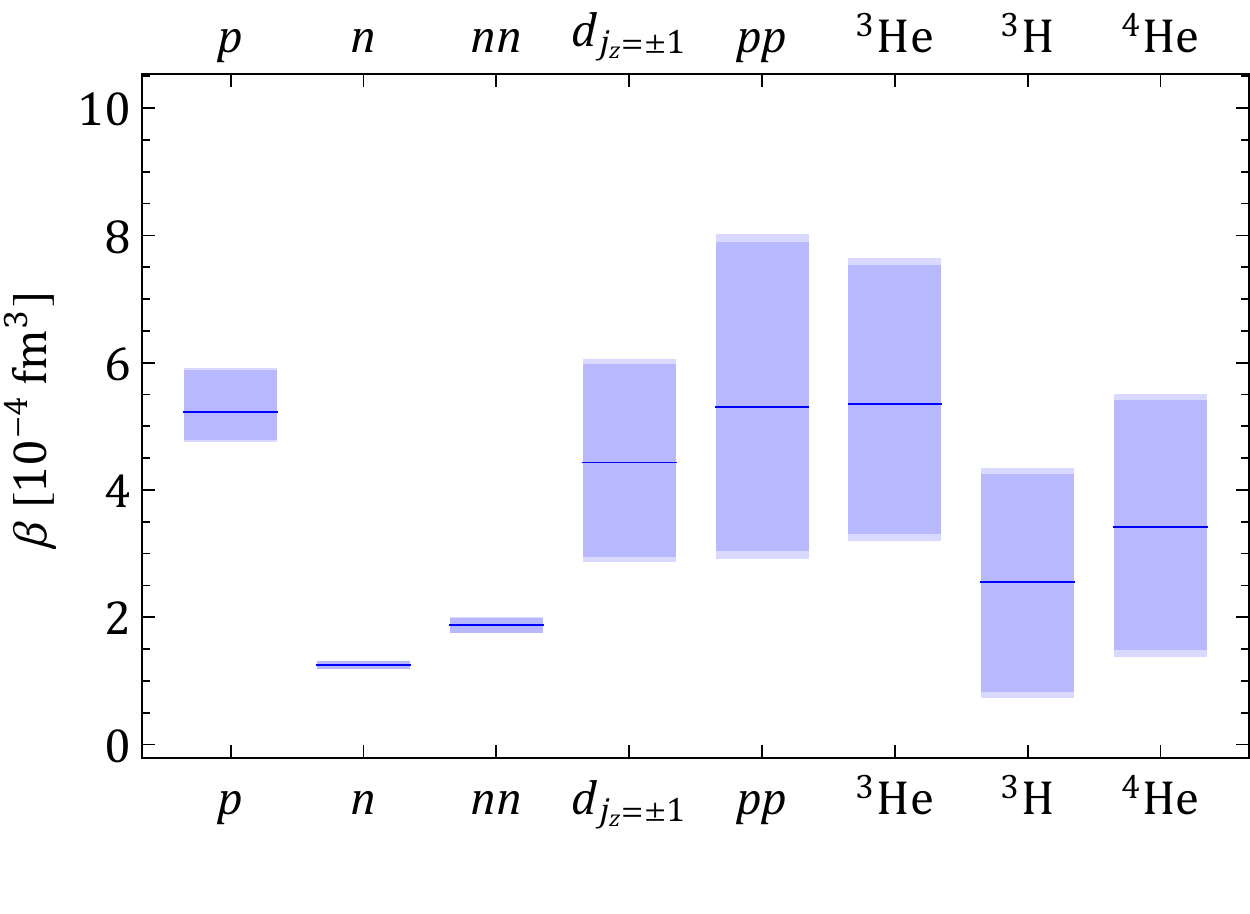}
 \vskip -0.15in
  \caption{
  The magnetic polarizabilities of the lightest nuclei at a pion mass of 
  $m_\pi\sim 805~{\rm MeV}$~\protect\cite{Chang:2015qxa}.  
         }
\label{fig:beta}
\end{figure}

In addition to nuclear magnetic moments, the magnetic polarizabilities have been calculated at 
$m_\pi\sim 805~{\rm MeV}$~\cite{Chang:2015qxa}, as shown in Fig.~\ref{fig:beta}.
The nucleon isovector polarizability is found to be substantial, in contrast to what is found in nature.
However, recent EFT calculations~\cite{Harald:2015CD}  
have shown that the cancellation between the pion-loop contributions 
and the $\Delta$-pole contribution weakens as the light-quark masses increase, leading to an increasing isovector component.

A few years ago, the first LQCD calculation of hadronic parity violation (HPV) was 
performed~\cite{Wasem:2011zz}.
This exploratory calculation
produced the interesting result of a small isovector conponent, which is consistent with experimental observation.
Given the ongoing experimental effort to observe HPV is $np\rightarrow d\gamma$~\cite{Gericke:2011zz},
there is some value in refining these calculations, both on the LQCD side and on the formal operator renormalization side.


Finally, we come to exotica in the form of penta-quarks, octa-quarks and beyond. 
Despite being exciting and a strong motivation for this work, 
I am not going to dwell on the fact that  LHCb has announced discovery of pentaquark systems 
of the form of  a $c\overline{c}$-$N$ resonant state~\cite{Aaij:2015tga}.
Since the early 1990's, nuclear physicists have been making efforts to quantify the interaction 
between quarkonia and nuclei.  
These are particularly interesting interactions as they are through the exchange of gluons only at leading order,
and therefore probe interactions that differ in nature from the nucleon-nucleon interaction.
As first discussed by Brodsky, Schmidt and de Teramond, structure in the pp 
spin-spin correlation just below the $c\overline{c}$ threshold suggests strong interactions 
between charmonia and the nucleon~ \cite{Brodsky:1989jd,Wasson:1991fb}.
After all this time, there is now an experimental program at JLab~\cite{Athenna}
 to look for the interactions between charmonia and the nucleon or light nuclei permitted by the 12 GeV upgrade.
\begin{figure}[!ht]
  \centering  
 \includegraphics[height=2.0 in ]{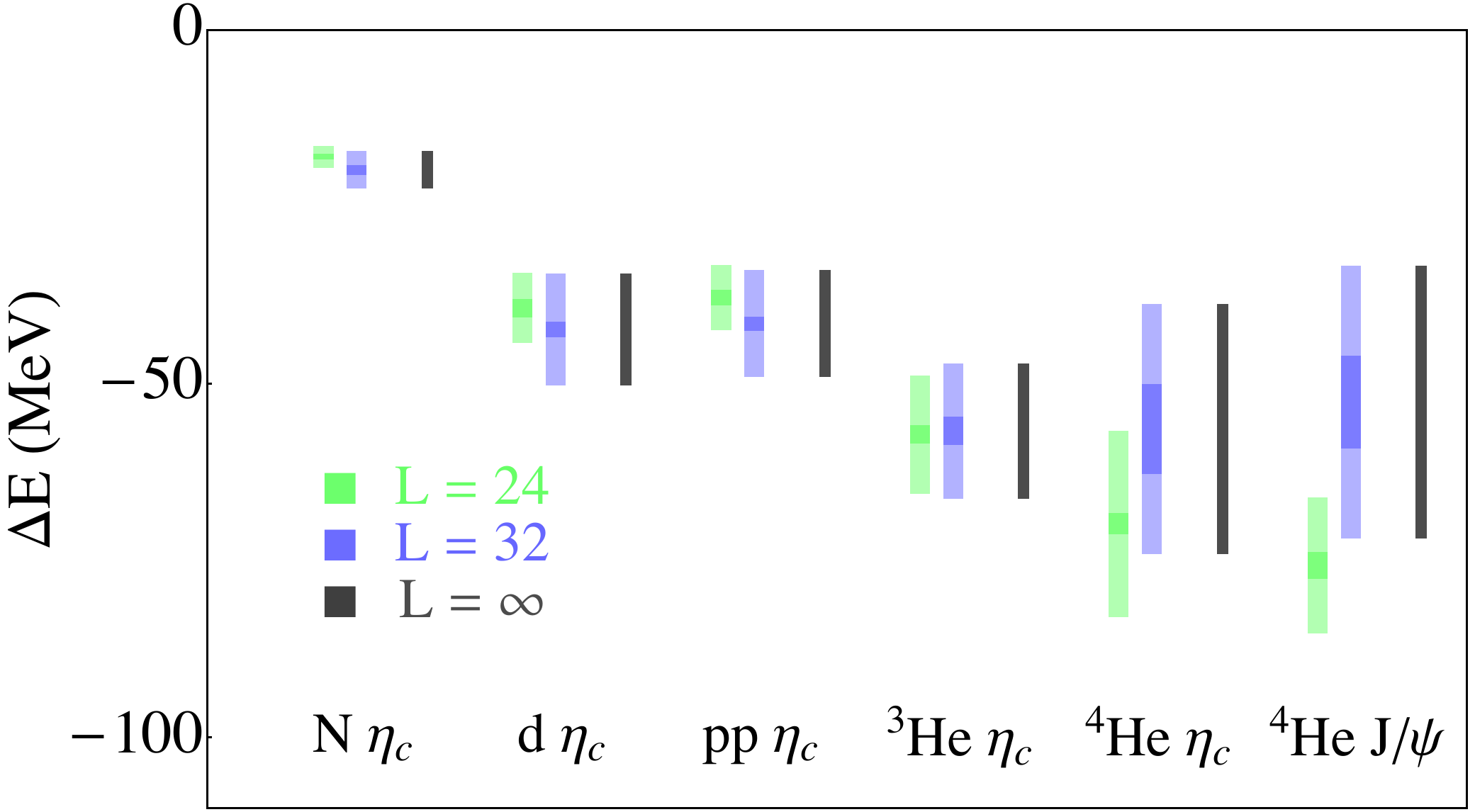}
 \vskip -0.15in
  \caption{
  The  calculated binding energies of charmonia to light nuclei~\protect\cite{Beane:2014sda}.
          }
\label{fig:Quarkonia}
\end{figure}
In anticipation of results from Jlab,
we have recently performed the first LQCD calculations of the interactions between charmonia and light nuclei
(at a heavy pion mass)~\cite{Beane:2014sda}, as shown in Fig.~\ref{fig:Quarkonia}.  
Assuming that the  A=4 binding energy is  the same as that to nuclear matter, 
 a binding energy of $\sim 40~{\rm MeV}$ is estimated in nuclear matter
at the physical light-quark masses,
determined from  the leading
behavior of the interaction.

\section{Summary}

It is interesting to consider what has happened during the period of the last nuclear physics long range plan (LRP), 
2007-2015, and what can be expected during the next one, 2015-2022.

During the last LRP, LQCD calculations of the structure of the nucleon evolved from either being quenched, $n_f=0$,  
or at relatively heavy quark masses, to fully dynamical light-quarks, $n_f=2+1$,
near and  at the physical values.  Further, dynamical QED is now being included in some calculations.
Calculations are being performed in multiple lattice volumes, at multiple lattice spacings 
(including relatively small spacings) and with multiple discretizations,  
permitting a complete quantification of uncertainties.
During this period, calculations related to the spectroscopy of hadrons have made  similarly impressive  advances, as previously highlighted.  The challenges facing calculations of these qualities are somewhat different to those of  structure calculations.  
In structure calculations, it is generally the properties of the ground state that are sought, while for spectroscopy, it is the location and content of many states above the ground state, which requires technology to cleanly isolate multiple exponentials from (many) correlation functions.  Resolution at short times is critical to being able to accomplish this, and hence the need for small lattice spacings in the time direction. 
This challenge, and the fact that the formalism is only now being put into place to rigorously extract S-matrix elements of complex states, has made it challenging to push toward the physical light-quark masses.
Typical calculations are now being performed in multiple lattice volumes with high precision, but at one lattice spacing only.
The spectra and interactions of multi-nucleon/hyperon systems face further distinct challenges.
The exponential degradation of the signal-to-noise is a challenge, and while mitigated by calculations at intermediate times, still makes calculations in even $A=3$ systems challenging.
Calculations have been comfortably performed at heavy pion masses in multiple lattice volumes, 
as discussed previously, but only at one lattice spacing.

\begin{figure}[!ht]
  \centering  
 \includegraphics[height=2.0 in ]{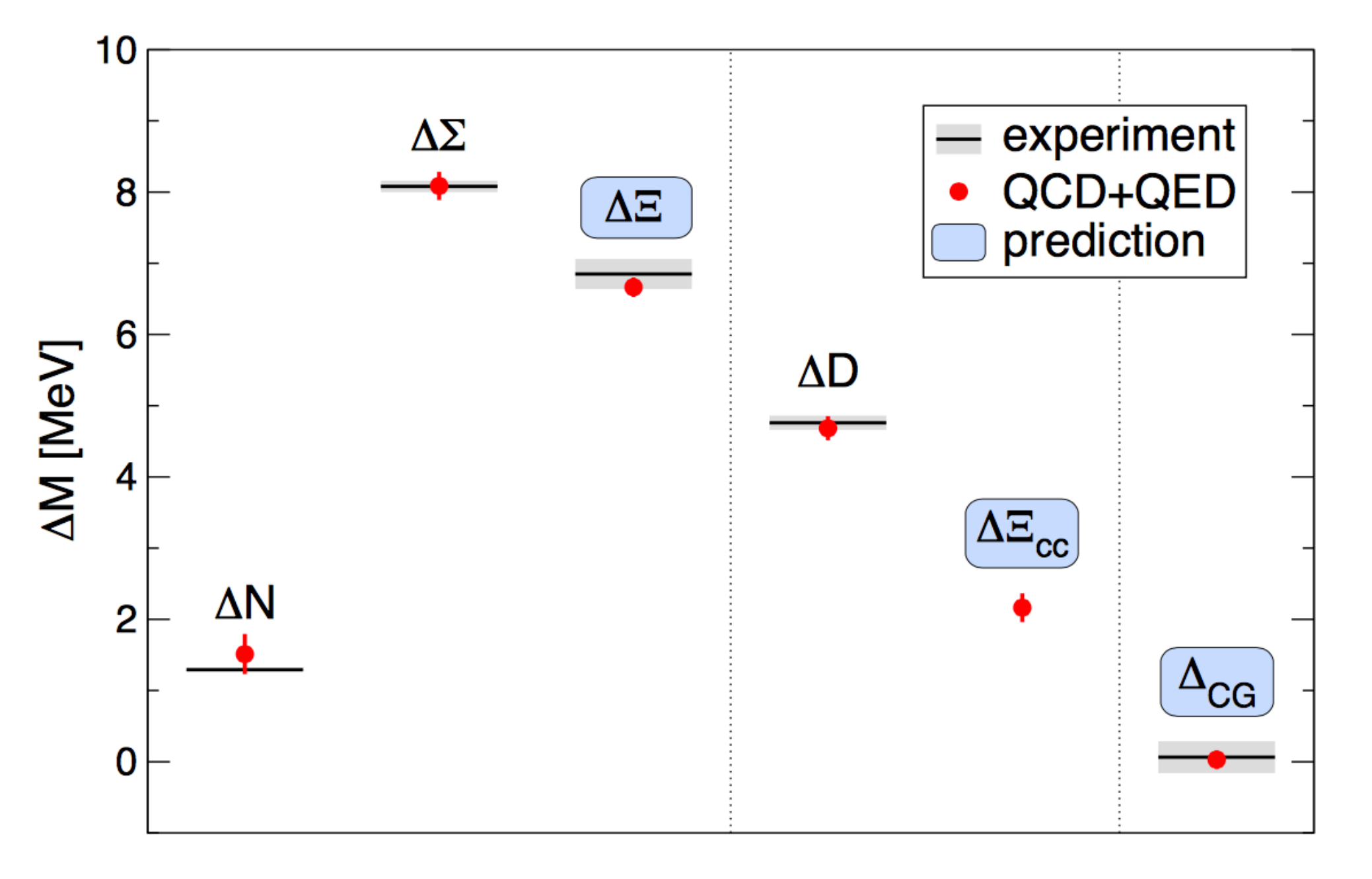}
 \vskip -0.15in
  \caption{
  Isospin-breaking mass differences calculated by the BMW collaboration~\cite{Borsanyi:2014jba} induced by the differences between the up-quark and down-quark masses and by fully-dynamical QED.
  [Figure reproduced with permission from Zoltan Fodor.]
          }
\label{fig:BMWisospin}
\end{figure}
If the computational resources grow as projected,
with the deployment of pre-exascale computers in 2017 and exascale computers a few years later, 
 and support for algorithm 
and code development/evolution continues to grow, the next period is going to be truly remarkable for 
nuclear physics. 
We should  expect to see typical LQCD calculations performed at the physical values of the light-quark masses, 
$n_f=1+1+1+1$
(includes dynamical charm quarks), with the inclusion of dynamical QED.
The calculations will be performed in multiple lattice volumes, with multiple lattice spacings and discretizations,
 and with sufficiently high precision to permit a complete quantification of associated uncertainties.
They will provide crucial input to other theoretical programs, such as refining the chiral nuclear forces for  nuclear many-body calculations.
Further, they will complement and  guide
the extensive experimental program of the field.
Part of my reason for optimism is recent work by the BMW collaboration~\cite{Borsanyi:2014jba}, 
and others, for example Refs.~\cite{Blum:2010ym,Zhou:2014gga,Horsley:2015eaa},
in 
which they calculate the isospin breaking between the the octet baryons and other quantities 
sensitive to isospin breaking and electromagnetism, the results of which are shown in Fig.~\ref{fig:BMWisospin}.
Particularly impressive is the prediction of the neutron-proton mass difference and its agreement 
with that found in nature.  
This bodes  well for the  future of nuclear physics.

Lattice QCD is 
on the brink of providing 
 first principles predictive capabilities for nuclear physics.
Calculations are  beginning to be performed at the physical light-quark masses and 
QED is now being included for some quantities.
With anticipated increases in computational resources and support for a HPC-trained workforce,
the next few years is going to see remarkable progress in directly connecting QCD to nuclear physics and in our ability to reliably predict important observables with fully-quantified uncertainties.

\Acknowledgements
I am grateful to the NPLQCD collaboration, and also to the USQCD collaboration,  for many discussions surrounding the physics 
 in this presentation.

\end{document}

%% file: econfmacros.tex
\def\beq{\begin{equation}}
\def\eeq#1{\label{#1}\end{equation}}
\def\eeqn{\end{equation}}

\def\beqa{\begin{eqnarray}}
\def\eeqa#1{\label{#1}\end{eqnarray}}
\def\eeqan{\end{eqnarray}}

\let\bar=\overbar

\def\Dslash{\not{\hbox{\kern-4pt $D$}}}
\def\dslash{\not{\hbox{\kern-2pt $\del$}}}

\def\msb{{\bar{\ssstyle M \kern -1pt S}}}




%% file: Savage_CIPANP_2015_v1p2.bbl
\begin{thebibliography}{99}

\bibitem{Fritzsch:1973pi} 
  H.~Fritzsch, M.~Gell-Mann and H.~Leutwyler,
  Phys.\ Lett.\ B {\bf 47}, 365 (1973).
  
\bibitem{Politzer:1973fx} 
  H.~D.~Politzer,
  Phys.\ Rev.\ Lett.\  {\bf 30}, 1346 (1973).
  
\bibitem{Politzer:1974fr} 
  H.~D.~Politzer,
  Phys.\ Rept.\  {\bf 14}, 129 (1974).
  
\bibitem{Gross:1973id} 
  D.~J.~Gross and F.~Wilczek,
  Phys.\ Rev.\ Lett.\  {\bf 30}, 1343 (1973).
  
\bibitem{Wilson:1974sk} 
  K.~G.~Wilson,
  Phys.\ Rev.\ D {\bf 10}, 2445 (1974).
  
  
  \bibitem{usqcd}
  USQCD Collaboration. 
 {\tt  http://usqcd.fnal.gov }
   
  
 \bibitem{scidac}
 Office of Science, Scientific Discovery through Advanced Computing, 
 {\tt http://www.scidac.gov }
  
  
\bibitem{Kronfeld:2012uk} 
  A.~S.~Kronfeld,
  Ann.\ Rev.\ Nucl.\ Part.\ Sci.\  {\bf 62}, 265 (2012)
  [arXiv:1203.1204 [hep-lat]].
  
  
  
  
\bibitem{gluex}
GlueX experiment,
{\tt http://www.gluex.org/GlueX/Home.html }
  
  
  
\bibitem{Wilson:2015dqa} 
  D.~J.~Wilson, R.~A.~Briceno, J.~J.~Dudek, R.~G.~Edwards and C.~E.~Thomas,
  arXiv:1507.02599 [hep-ph].
  
\bibitem{Wilson:2014cna} 
  D.~J.~Wilson, J.~J.~Dudek, R.~G.~Edwards and C.~E.~Thomas,
  Phys.\ Rev.\ D {\bf 91}, no. 5, 054008 (2015)
  [arXiv:1411.2004 [hep-ph]].
  
\bibitem{Morningstar:2013bda} 
  C.~Morningstar, J.~Bulava, B.~Fahy, J.~Foley, Y.~C.~Jhang, K.~J.~Juge, D.~Lenkner and C.~H.~Wong,
  Phys.\ Rev.\ D {\bf 88}, no. 1, 014511 (2013)
  [arXiv:1303.6816 [hep-lat]].
  
\bibitem{Briceno:2012yi} 
  R.~A.~Briceno and Z.~Davoudi,
  Phys.\ Rev.\ D {\bf 88}, no. 9, 094507 (2013)
  [arXiv:1204.1110 [hep-lat]].
  
  
  
  
\bibitem{Lee:2015jqa} 
  G.~Lee, J.~R.~Arrington and R.~J.~Hill,
  Phys.\ Rev.\ D {\bf 92}, no. 1, 013013 (2015)
  [arXiv:1505.01489 [hep-ph]].
  
  
  
  
  
\bibitem{Shanahan:2014tja} 
  P.~E.~Shanahan {\it et al.},
  Phys.\ Rev.\ Lett.\  {\bf 114}, no. 9, 091802 (2015)
  [arXiv:1403.6537 [hep-lat]].
  
\bibitem{Bhattacharya:2011qm} 
  T.~Bhattacharya, V.~Cirigliano, S.~D.~Cohen, A.~Filipuzzi, M.~Gonzalez-Alonso, M.~L.~Graesser, R.~Gupta and H.~W.~Lin,
  Phys.\ Rev.\ D {\bf 85}, 054512 (2012)
  [arXiv:1110.6448 [hep-ph]].
  
\bibitem{Gupta:2014dla} 
  R.~Gupta, T.~Bhattacharya, A.~Joseph, S.~D.~Cohen and H.~W.~Lin,
  PoS LATTICE {\bf 2013}, 409 (2014)
  [arXiv:1403.2447 [hep-lat]].
  
\bibitem{Bhattacharya:2015wna} 
  T.~Bhattacharya, V.~Cirigliano, S.~Cohen, R.~Gupta, A.~Joseph, H.~W.~Lin and B.~Yoon,
  arXiv:1506.06411 [hep-lat].
  
  
  
\bibitem{Weinberg:1990rz} 
  S.~Weinberg,
  Phys.\ Lett.\ B {\bf 251}, 288 (1990).
  
\bibitem{Weinberg:1991um} 
  S.~Weinberg,
  Nucl.\ Phys.\ B {\bf 363}, 3 (1991).
  
  
\bibitem{Ordonez:1992xp} 
  C.~Ordonez and U.~van Kolck,
  Phys.\ Lett.\ B {\bf 291}, 459 (1992).
  
\bibitem{Kaplan:1996xu} 
  D.~B.~Kaplan, M.~J.~Savage and M.~B.~Wise,
  Nucl.\ Phys.\ B {\bf 478}, 629 (1996)
  [nucl-th/9605002].
  
\bibitem{Kaplan:1998tg} 
  D.~B.~Kaplan, M.~J.~Savage and M.~B.~Wise,
  Phys.\ Lett.\ B {\bf 424}, 390 (1998)
  [nucl-th/9801034].
  
\bibitem{Kaplan:1998we} 
  D.~B.~Kaplan, M.~J.~Savage and M.~B.~Wise,
  Nucl.\ Phys.\ B {\bf 534}, 329 (1998)
  [nucl-th/9802075].
  
\bibitem{Beane:2001bc} 
  S.~R.~Beane, P.~F.~Bedaque, M.~J.~Savage and U.~van Kolck,
  Nucl.\ Phys.\ A {\bf 700}, 377 (2002)
  [nucl-th/0104030].
  
\bibitem{Epelbaum:2008ga} 
  E.~Epelbaum, H.~W.~Hammer and U.~G.~Meissner,
  Rev.\ Mod.\ Phys.\  {\bf 81}, 1773 (2009)
  [arXiv:0811.1338 [nucl-th]].
  
\bibitem{Machleidt:2011zz} 
  R.~Machleidt and D.~R.~Entem,
  Phys.\ Rept.\  {\bf 503}, 1 (2011)
  [arXiv:1105.2919 [nucl-th]].
  
\bibitem{Beane:2012vq} 
  S.~R.~Beane {\it et al.} [NPLQCD Collaboration],
  Phys.\ Rev.\ D {\bf 87}, no. 3, 034506 (2013)
  [arXiv:1206.5219 [hep-lat]].
  
\bibitem{Beane:2013br} 
  S.~R.~Beane {\it et al.} [NPLQCD Collaboration],
  Phys.\ Rev.\ C {\bf 88}, no. 2, 024003 (2013)
  [arXiv:1301.5790 [hep-lat]].
  
\bibitem{Beane:2011iw} 
  S.~R.~Beane {\it et al.} [NPLQCD Collaboration],
  Phys.\ Rev.\ D {\bf 85}, 054511 (2012)
  [arXiv:1109.2889 [hep-lat]].
  
\bibitem{Yamazaki:2012hi} 
  T.~Yamazaki, K.~i.~Ishikawa, Y.~Kuramashi and A.~Ukawa,
  Phys.\ Rev.\ D {\bf 86}, 074514 (2012)
  [arXiv:1207.4277 [hep-lat]].
  
\bibitem{Yamazaki:2015asa} 
  T.~Yamazaki, K.~i.~Ishikawa, Y.~Kuramashi and A.~Ukawa,
  Phys.\ Rev.\ D {\bf 92}, no. 1, 014501 (2015)
  [arXiv:1502.04182 [hep-lat]].
  
\bibitem{Orginos:2015aya} 
  K.~Orginos, A.~Parre\~no, M.~J.~Savage, S.~R.~Beane, E.~Chang and W.~Detmold,
  arXiv:1508.07583 [hep-lat].
  
\bibitem{Beane:2006mx} 
  S.~R.~Beane, P.~F.~Bedaque, K.~Orginos and M.~J.~Savage,
  Phys.\ Rev.\ Lett.\  {\bf 97}, 012001 (2006)
  [hep-lat/0602010].
  
  
  
\bibitem{Barnea:2013uqa} 
  N.~Barnea, L.~Contessi, D.~Gazit, F.~Pederiva and U.~van Kolck,
  Phys.\ Rev.\ Lett.\  {\bf 114}, no. 5, 052501 (2015)
  [arXiv:1311.4966 [nucl-th]].
  
  
\bibitem{vanKolck:1998bw} 
  U.~van Kolck,
  Nucl.\ Phys.\ A {\bf 645}, 273 (1999)
  [nucl-th/9808007].
  
\bibitem{Chen:1999tn} 
  J.~W.~Chen, G.~Rupak and M.~J.~Savage,
  Nucl.\ Phys.\ A {\bf 653}, 386 (1999)
  [nucl-th/9902056].
  
 
 
 
\bibitem{Beane:2013kca} 
  S.~R.~Beane, S.~D.~Cohen, W.~Detmold, H.-W.~Lin and M.~J.~Savage,
  Phys.\ Rev.\ D {\bf 89}, 074505 (2014)
  [arXiv:1306.6939 [hep-ph]].
  
  
  
\bibitem{Beane:2014ora} 
  S.~R.~Beane {\it et al.},
  Phys.\ Rev.\ Lett.\  {\bf 113}, no. 25, 252001 (2014)
  [arXiv:1409.3556 [hep-lat]].
   
\bibitem{Chang:2015qxa} 
  E.~Chang, W.~Detmold, K.~Orginos, A.~Parre\~no, M.~J.~Savage, B.~C.~Tiburzi and S.~R.~Beane,
  arXiv:1506.05518 [hep-lat].
  
\bibitem{Beane:2015yha} 
  S.~R.~Beane, E.~Chang, W.~Detmold, K.~Orginos, A.~Parre\~no, M.~J.~Savage and B.~C.~Tiburzi,
  arXiv:1505.02422 [hep-lat].
  
  
\bibitem{Harald:2015CD} 
H.~Griesshammer,
talk presented at
{\bf Chiral Dynamics 2015},
June (2015), Pisa, Italy.
  
  
\bibitem{Wasem:2011zz} 
  J.~Wasem,
  Phys.\ Rev.\ C {\bf 85}, 022501 (2012)
  [arXiv:1108.1151 [hep-lat]].
 
\bibitem{Gericke:2011zz} 
  M.~T.~Gericke {\it et al.},
  Phys.\ Rev.\ C {\bf 83}, 015505 (2011).
   
   
\bibitem{Aaij:2015tga} 
  R.~Aaij {\it et al.} [LHCb Collaboration],
  Phys.\ Rev.\ Lett.\  {\bf 115}, 072001 (2015)
  [arXiv:1507.03414 [hep-ex]].
  
  
\bibitem{Brodsky:1989jd} 
  S.~J.~Brodsky, I.~A.~Schmidt and G.~F.~de Teramond,
  Phys.\ Rev.\ Lett.\  {\bf 64}, 1011 (1990).
  
\bibitem{Wasson:1991fb} 
  D.~A.~Wasson,
  Phys.\ Rev.\ Lett.\  {\bf 67}, 2237 (1991).
   
 \bibitem{Athenna}
 {\tt  http://www.jlab.org/exp\_prog/proposals/12/PR12-12-006.pdf}
  
\bibitem{Beane:2014sda} 
  S.~R.~Beane, E.~Chang, S.~D.~Cohen, W.~Detmold, H.-W.~Lin, K.~Orginos, A.~Parre\~no and M.~J.~Savage,
  Phys.\ Rev.\ D {\bf 91}, no. 11, 114503 (2015)
  [arXiv:1410.7069 [hep-lat]].
  
\bibitem{Borsanyi:2014jba} 
  S.~Borsanyi {\it et al.},
  Science {\bf 347}, 1452 (2015)
  [arXiv:1406.4088 [hep-lat]].
  
\bibitem{Blum:2010ym} 
  T.~Blum, R.~Zhou, T.~Doi, M.~Hayakawa, T.~Izubuchi, S.~Uno and N.~Yamada,
  Phys.\ Rev.\ D {\bf 82}, 094508 (2010)
  [arXiv:1006.1311 [hep-lat]].
  
\bibitem{Zhou:2014gga} 
  R.~Zhou {\it et al.} [MILC Collaboration],
  PoS LATTICE {\bf 2014}, 024 (2014)
  [arXiv:1411.4115 [hep-lat]].
  
 
\bibitem{Horsley:2015eaa} 
  R.~Horsley {\it et al.},
  arXiv:1508.06401 [hep-lat].
  
  
  
\end{thebibliography}
